\documentclass[a4paper,10pt]{iopart}
 \usepackage{cases}

\usepackage{iopart}
\usepackage{epsfig}
\usepackage{cite}
\usepackage{setspace}

\begin{document}
\title[Competition between MI and SBS in a Broadband Slow Light Device]
{Competition between the Modulation Instability and Stimulated
Brillouin Scattering in a Broadband Slow Light Device}

\author{
Yunhui Zhu,$^{1,*}$  E. Cabrera-Granado,$^1$ Oscar G. Calderon,$^2$\\
Sonia Melle,$^2$  Yoshitomo Okawachi,$^3$ Alexander L. Gaeta,$^3$
\\and Daniel J. Gauthier$^1$}
\address{$^1$Department of Physics and the Fitzpatrick Institute for Photonics, Duke University, Durham,
North Carolina, 27708 USA.}
\address{ $^2$Escuela Universitaria de
Optica, Universidad Complutense de Madrid, C/Arcos de Jal\'{o}n s/n,
28037 Madrid, Spain}
\address{ $^3$School of Applied and Engineering
Physics, Cornell University, Ithaca, NY, 14853, USA}
\ead{$^*$yz65@phy.duke.edu}

\begin{abstract}
We observe competition between the modulation instability (MI) and
stimulated Brillouin scattering (SBS) in a 9.2-GHz broadband SBS
slow light device, in which a standard 20-km-long single-mode LEAF
fibre is used as the SBS medium. We find that MI is dominant and
depletes most of the pump power when we use an intense pump beam at
$\sim$1.55 $\mu$m, where the LEAF fibre is anomalously dispersive.
The dominance of the MI in the LEAF-fibre-based system suppresses
the SBS gain, degrading the SBS slow light delay and limiting the
SBS gain-bandwidth to 126 dB$\cdot$GHz. In a dispersion-shifted
highly nonlinear fibre, the SBS slow light delay is improved due to
the suppression of the MI, resulting in a gain-bandwidth product of
344 dB$\cdot$GHz, limited by our available pump power of 0.82~W.
\end{abstract}
 \submitto{\JO}

\maketitle

\section{Introduction}
Slow light refers to the dramatic reduction of the group velocity
$v_g$ for optical pulses propagating through a dispersive material
\cite{controlling}. Among various slow light approaches,
stimulated-Brillouin-scattering-based slow light in single-mode
optical fibres has attracted much interest. Because of the optically
controllable delay time and tunability of the bandwidth
\cite{controlSBSslow,tunableSBSslow}, stimulated Brillouin
scattering (SBS) slow light devices have great potential for
all-optical applications such as network buffering, optical packet
switching, and data synchronization \cite{buffer,synchronization}.
Fractional delays as large as 3 have been demonstrated recently
\cite{broadbandEdu-best}.

In SBS slow light systems, the bandwidth of the device is determined
by the linewidth of the resonance. The intrinsic linewidth
$\Gamma_B$ of the resonance in normal single-mode fibre is 40 MHz
(FWHM), which is determined by the decay rate of acoustic phonons in
the optical fibre. The data rate in such a system is limited to
$\sim$$\Gamma_B$, which is insufficient for modern optical
communication applications. To solve this problem, broadband SBS
slow light has been developed, in which the linewidth of the SBS
resonance is broadened using a multi-frequency laser source as the
pump beam \cite{broadband-two,shi2007design}.

Herr{\'a}ez \textit{et al.} \cite{broadband1st-continous} first
demonstrated tailoring of the pump spectrum using direct modulation
of the injecting current of the pump laser. Since then, a number of
research groups have successfully broadened the bandwidth of the SBS
slow light up to tens of GHz, which is much larger than the
intrinsic linewidth of the SBS resonance
\cite{song200725,yi2007,broadbandEdu-best}. However, as we broaden
the pump beam spectrum, its power spectral density decreases,
thereby suppressing the SBS gain. As a result, the slow light delay,
which is proportional to the SBS gain, also decreases. Therefore,
the power of the pump beam must be increased to compensate for the
reduction of gain at larger bandwidths \cite{broadband-gain}. As the
power increases, other nonlinear optical processes can become
important if their response time is fast in comparison to the
inverse of the pump beam spectral width. For example, stimulated
Raman scattering (SRS) and the modulation instability (MI) have
response times of $\sim$1~ps\cite{stolen1989raman} and
$\sim$10~fs\cite{NonlinearFiberOptics} respectively, and hence are
not affected by pump broadening of several GHz. Eventually, as the
bandwidth of the pump is increased and the power increased in
proportion, SRS and/or the MI becomes more efficient, potentially
robbing pump power and limiting the SBS gain.

Olsson \textit{et al.} \cite{olsson1987} predict that SRS dominates
over SBS for pump bandwidths larger than 20 GHz for a gain of 20 dB.
No experiment validation of this prediction has been made to the
best of our knowledge. On the contrary, our experiment shows that
the MI is the dominant competing process in a 9.2-GHz-bandwidth SBS
slow light device, which limits the SBS gain-bandwidth product.

In this paper, we study the competing effects and the related
limitations on broadband SBS slow light. First, we introduce our
broadband SBS slow light experiment and describe the competing
processes that can possibly suppress the SBS process. We then verify
that SBS is suppressed due to the MI in our broadband SBS device,
where a standard single-mode LEAF fibre is used as the SBS medium.
We focus here on the LEAF fibre because it is a readily available,
relatively inexpensive fibre used in long-haul communication systems
and hence is likely to be selected for SBS slow light devices.  We
also use a wavelength of 1.55 $\mu$m because it is in the centre of
the telecommunication band. As a solution to the problem, we show
that the MI and the associated SBS slow light degradation are
eliminated by using a dispersion shifted fibre that has normal
dispersion at the 1.55 $\mu$m wavelength window. Finally, we compare
the SBS gain-bandwidth-product limit imposed by the MI with the
limit imposed by SRS following the approach of Olsson \textit{et
al.}\cite{olsson1987} in the LEAF-based system.

\section{Broadband SBS slow light and the competing processes }

In a broadband SBS slow light device, an optical pulse (the signal)
is delayed by interacting with a counterpropagating broadband pump
beam via the SBS process. As shown in figure 1, two beams (signal
and pump) counterpropagate through the slow-light medium (fibre),
where they interact and create low-frequency acoustic waves via
electrostriction. These acoustic phonons, in turn, scatter the
optical beams via Brillouin scattering. When the frequency of the
signal beam is tuned to the Stokes line, that is, downshifted by the
Brillouin frequency $\Omega_B$ from the frequency of the pump beam,
the opto-acoustic coupling becomes strong, and light from the pump
beam is efficiently scattered into the signal beam, inducing a gain
resonance and giving rise to a variation in the refractive index in
a narrow frequency range around the resonance frequency, which
results in a small $v_g$ for the signal beam \cite{NonlinearOptics}.

\begin{figure}[hb]\label{setup}
\flushright\includegraphics[width=10cm]{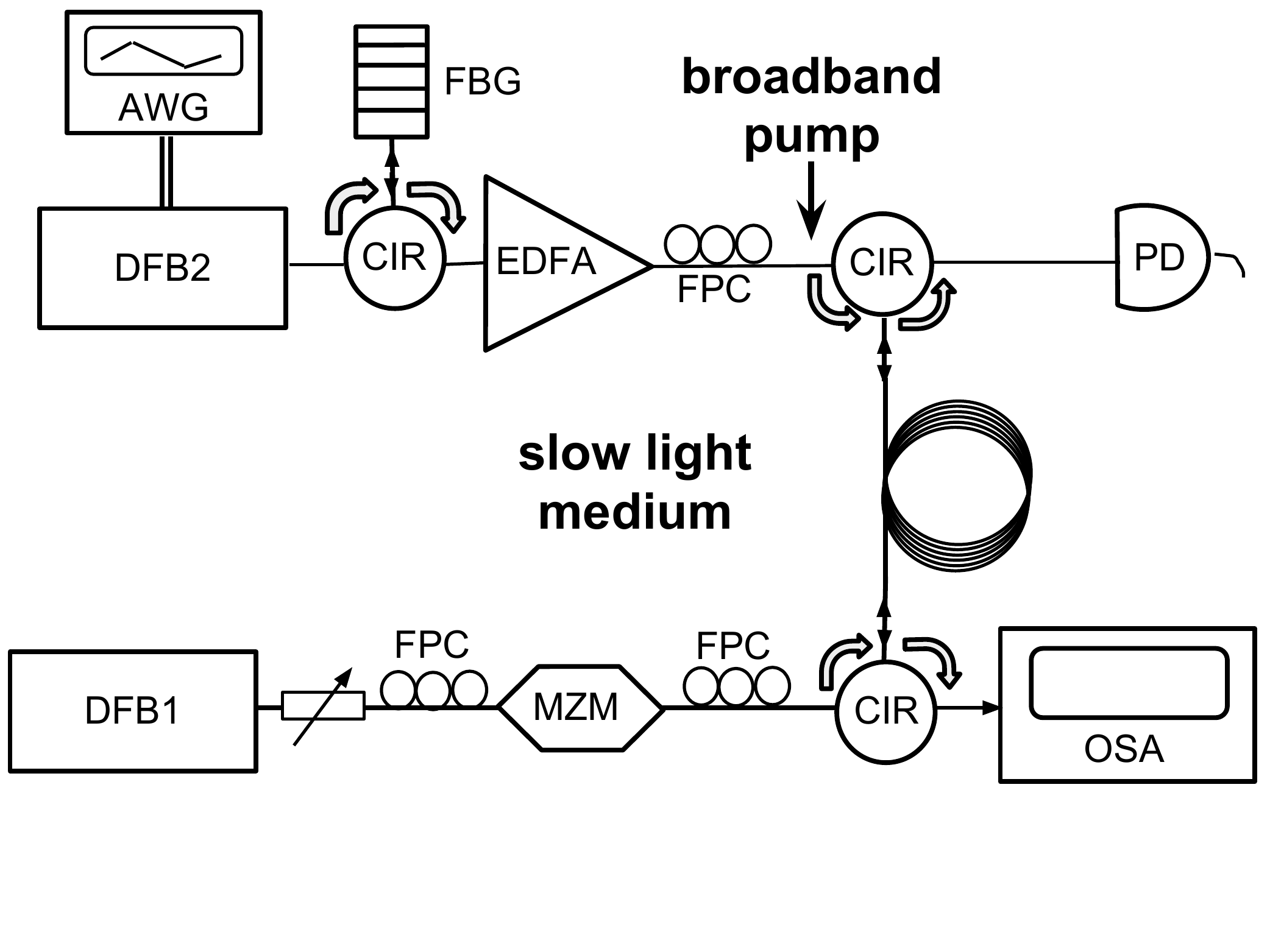} \caption{Experimental
Setup. The injection current of the pump laser (DFB2) is modulated
by an arbitrary wavefunction generator (AWG). The current modulation
function is tailored to produce a flat-topped broadened gain profile
\cite{broadbandEdu-best}. A Fibre Bragg Grating (FBG) (bandwidth
$\Delta\nu=24$ GHz, central wavelength $\sim$1.55 $\mu$m) is used to
filter out unwanted frequency sidebands from the DFB laser. An EDFA
amplifies the pump beam before it is injected into the fibre via a
circulator. The signal beam (DFB1) is modulated with a Mach-Zehnder
modulator (MZM) and is injected from the other end of the fibre via
a circulator. Fibre polarization controllers (FPC) are used to match
the polarization orientations of the pump and signal beams. A
photodiode (PD) measures the output power of the signal beam and an
optical spectrum analyzer (OSA) measures the spectrum of the output
pump beam. }
\end{figure}

Consider an optical pulse propagating through a fibre of length $L$,
the transit time $T_g$ is approximately given by
\cite{boydmaximum,SBStheory}
\begin{equation}
T_g=\frac{L}{v_{g}}=\frac{Ln_{g}}{c},
\end{equation}
where
\begin{equation}\label{group_index}
n_{g}=n+\nu\frac{\rmd n}{\rmd\nu}
\end{equation}
is the group index, $n$ is the refractive index, $\nu$ is the
frequency (in cycles/s) and $c$ is the speed of light in vacuum. The
delay time $T_{d}$ in a slow light system is defined as the
difference in the transit time through the medium with and without
the slow light effect, namely
\begin{equation}\label{delaydefine}
    T_{d}=T_{g}-T_{g0}=\frac{L}{c}(n_{g}-n_{g0}),
\end{equation}
where $n_g$ and $n_{g0}$ are the group indices with and without slow
light effect, respectively.

 In our broadband SBS slow light experiment, a
current-modulated distributed feedback laser (DFB) is used to
increase the linewidth of the SBS resonance. By designing the
modulation function, we obtain a square-shaped pump-beam spectrum
\cite{broadbandEdu-best}. The broadened complex SBS gain profile
$\tilde{g}(\nu)$ results from the convolution of the intrinsic
complex SBS gain spectrum $\tilde{g_0}(\nu)$ with the intensity
spectrum $i_{i}(\nu_{p})$ of the input pump beam, which is given by
\cite{SBStheory,broadband-gain}
\begin{eqnarray}\label{convolution}
     \tilde{g}(\nu)&=\tilde{g_0}(\nu)\otimes i_{i}(\nu_{p})  \\
     &=\int_{-\infty}^{\infty}\frac{(g_0P_{i}/A_{eff}\Gamma)\cdot
     \textrm{rect}[(\nu_{p}-\nu_{{p}0})/\Gamma]}{1-\rmi(\nu+\Omega_{B}-\nu_{p})/(\Gamma_{B}/2)}d\nu_{p},\label{flat}
\end{eqnarray}
where
\begin{subnumcases}
{\textrm{rect}(x)=}
1 &$|x|<1/2$\\
0 &$|x|>1/2$,
\end{subnumcases}
$g_0$ is the line centre SBS gain factor (a constant determined by
the material), $P_{i}=\int i_{i}(\nu_{p})d\nu_{p}\cdot A_{eff}$ is
the input power of the pump beam, $A_{eff}$ is the effective
cross-section area of the fibre, $\Gamma=9.2$ GHz is the bandwidth
of the pump spectrum and $\nu_{p0}$ is the central frequency of the
pump-beam spectrum. \Eref{convolution} is obtained in the so-called
undepleted pump approximation, where the pump beam power $P_i$ is
constant along the fibre.

The gain profile of the broadened SBS resonance is obtained from the
real part of the integral in \eref{flat} and is given by
\begin{eqnarray}\label{gain_profile}
g(\delta)&=\frac{\Gamma_{B}}{2\Gamma}\frac{g_0P_{i}}{A_{eff}}
[\arctan(\frac{\delta+\Gamma/2}{\Gamma_{B}/2})-\arctan(\frac{\delta-\Gamma/2}{\Gamma_{B}/2})]\\
&\approx \frac{\pi \Gamma_{B}}{2\Gamma}\frac{g_0P_{i}}{A_{eff}}
\textrm{rect}(\delta/\Gamma) \qquad \textrm{when}\
\Gamma\gg\Gamma_{B},
\end{eqnarray}
where $\delta=\nu+\Omega_{B}-\nu_{{p}0}$ is the signal beam detuning
from resonance. We see that the bandwidth of the resonance is
broadened to $\Gamma$. However, $g(\delta)$ scales inversely
proportional to the bandwidth $\Gamma$ and more input power $P_{i}$
is required to compensate for the loss in gain.

The refractive index  $n$ associated with the SBS process is
obtained from the imaginary part of \eref{flat} and is given by
\begin{equation}\label{refractive_index}
n(\delta)=n_0+\frac{c}{2\pi\nu}\frac{\Gamma_{B}}{2\Gamma}\frac{g_0P_{i}}{A_{eff}}
\ln\frac{1+[(\delta+\Gamma/2)/(\Gamma_{B}/2)]^2}{1+[(\delta-\Gamma/2)/(\Gamma_{B}/2)]^2},
\end{equation}
where $n_0$ is the refractive index without the SBS process. The
group refractive index $n_{g}$ at zero detuning ($\delta=0$) is then
determined using \eref{group_index} and \eref{refractive_index} to
obtain
\begin{equation}\label{group refractive index}
    n_{g}=n+\nu\frac{\rmd
    n}{\rmd\nu}|_{\delta=0}\approx n_{g0}+\frac{c}{\pi}
    \frac{\Gamma_{B}}{\Gamma^2}\frac{g_0P_{i}}{A_{eff}}.
\end{equation}
Because we assume that the pump power $P_i$ does not change over the
fibre length $L$, the delay time of an on-resonance signal pulse
whose spectral bandwidth is much smaller than that of the broadband
SBS resonance is given approximately by
\begin{equation}\label{delay}
   T_{d}=\frac{\Gamma_{B}}{\pi\Gamma^2}\frac{g_0P_{i}L}{A_{eff}}=\frac{G}{\pi^2\Gamma},
\end{equation}
where
\begin{eqnarray}\label{accurategain}
G=\ln\frac{P_{s}}{P_{s0}}=2g(0)L
=2\frac{\Gamma_B}{\Gamma}\frac{g_0P_{i}L}{A_{eff}}\arctan(\Gamma/\Gamma_B)\\\label{gain}
\approx \frac{\pi\Gamma_{B}}{\Gamma}\frac{g_0P_{i}L}{A_{eff}}\qquad
\textrm{when}\ \Gamma\gg\Gamma_{B},
\end{eqnarray}
is the power gain experienced by the probe beam, and $P_s$ and
$P_{s0}$ are the output powers of the signal beam with and without
the pump beam, respectively. If attenuation in the fibre is
considered, the effective length $L_{eff}=[1-\exp(-\alpha
L)]/\alpha$ is used in place of length $L$ in equations
\eref{accurategain} and \eref{gain}, where  $\alpha$ is the
attenuation coefficient. We see from \eref{delay} and \eref{gain}
that the delay time $T_d$ is proportional to the gain $G$, which is
proportional to the pump power $P_{i}$ and inversely proportional to
the pump spectral bandwidth $\Gamma$. As a result, we are able to
control the delay time $T_d$ by adjusting the pump power.

There are a number of effects that limit the SBS gain $G$, thereby
limiting the SBS slow light delay. One ultimate limitation on the
SBS slow light delay is SBS gain saturation caused by pump
depletion. It takes place when the Stokes beam becomes so strong
that most of the power contained in the pump beam is transferred to
it. Two different saturation processes are involved. One occurs when
the initial probe beam is strong; the amplified probe beam grows
quickly and depletes the pump beam even for moderate $G$.  The other
one takes place when $G$ approaches a threshold value $G_{{th}}$,
found to be $\sim$10 in single-mode fibres \cite{threshold-fiber},
where spontaneous Brillouin scattering from thermal fluctuations is
amplified and becomes sufficiently large, depleting the pump power
even in the absence of a probe beam. When the pump beam is depleted,
the undepleted pump assumption is no longer valid. As a result, $G$,
and thus the delay time $T_d$, no longer grow with increasing pump
power \cite{threshold-fiber,NonlinearOptics}.

Stimulated Raman scattering (SRS) is another competing nonlinear
effect in which a pump beam is scattered by high-frequency optical
phonons \cite{NonlinearFiberOptics}. As a result, the frequency at
which the SRS gain resonance occurs is downshifted by $\sim$13 THz
from the frequency of the pump beam in single-mode fibres. In the
presence of a strong pump beam, the Stokes beam initiated by
spontaneous Raman scattering is amplified exponentially via the SRS
process. When the SRS gain reaches a threshold value of $\sim$10 in
single-mode fibres \cite{NonlinearFiberOptics}, most of the pump
power is transferred to the Stokes beam.

The SRS gain with a monochromatic pump is typically two
orders-of-magnitude smaller than the SBS gain. However, as a result
of the fast response time of SRS in single-mode fibres ($<$1 ps)
\cite{stolen1989raman}, the Raman gain is not affected by the
spectral broadening of the pump beam up to the value of 9.2 GHz used
in our experiment, whereas the SBS gain is inversely proportional to
the bandwidth of the pump as shown in \eref{gain}. As the pump
bandwidth increases, the SBS gain will eventually  match the SRS
gain, setting the scale for competition between these processes.

MI is yet another process that can compete with the broadband SBS
process. MI results from the interplay between the nonlinear Kerr
effect and material dispersion. It is a four-wave-mixing process
where two copropagating photons of the same frequency are converted
into a frequency up-shifted and down-shifted photon pair
\cite{NonlinearFiberOptics,MIpulse}. As a result, the MI broadens
the spectrum of continuous wave (cw) or quasi-cw beams, even turning
a continuous wave beam into a train of pulses \cite{MIabnormal,
MIbroadening,MIpulse}. In the presence of a strong cw or quasi-cw
beam propagating through the fibre, noise components in the vicinity
of peaks of the MI gain experience exponential amplification, which
leads to the creation of two symmetric spectral side lobes
\cite{NonlinearFiberOptics,MIpulse}. The gain profile of the MI is
given by \cite{NonlinearFiberOptics}
\begin{equation}\label{gainMI}
 G_{MI}(\nu)=4\pi^2|\beta_2\nu|\sqrt{2\Omega_{peak}^2-\nu^2}L,
\end{equation}
where $\Omega_{peak}$ (in cycles/s) is the frequency shift at which
the maximum gain is obtained, and $\beta_2$ is the group velocity
dispersion (GVD) parameter. Here \cite{NonlinearFiberOptics},
\begin{equation}\label{peak}
    \Omega_{peak}=\pm\frac{1}{2\pi}\sqrt{\frac{2\gamma P_{i}}{|\beta_2|}},
\end{equation}
where $\gamma$ is the nonlinear parameter. Similar to SRS, the MI is
not affected by the pump broadening of 9.2 GHz due to its fast
response time ($<$10 fs). We will show that MI is the dominant
competing effect in our SBS experiment at high input pump power
$P_i$ in a LEAF fibre.

\section{Experiment results}

In the experiment, as shown in figure 1, a distributed feedback
(DFB) laser operating at $\sim$1.55 $\mu$m is used as the pump
source. We modulate the injecting current of the DFB laser with a
modified triangle function so that the output beam of the DFB laser
has a square-shaped spectrum with a bandwidth of 9.2 GHz
\cite{broadbandEdu-best}. The output of the DFB laser is then
amplified with an erbium-doped fibre amplifier (EDFA) to provide
enough pump power for the broadband SBS process. The EDFA also
controls the input pump power and thereby controls the SBS gain.
Another DFB laser is used to generate the signal beam, which is
tuned to the SBS resonance. To avoid probe-induced SBS gain
saturation, the signal beam is attenuated to a power of 2 $\mu$W
before it is injected into the SBS medium, where it
counterpropagates and interacts with the pump beam via the SBS
process. The amplified and delayed signal is detected at the output.
The gain $G$ is obtained by measuring the output powers of the
signal beam with the pump beam on and off.

We use two different fibres as the SBS medium in the experiment, a
20-km-long LEAF fibre from Corning and a 2-km-long HNLF fibre from
OFS. The parameters of the fibres are shown in Table 1. The GVD
parameter $\beta_2$ is measured by the time-of-flight method
\cite{TOF}, in which the group velocity as function of the
wavelength is measured by recording the transit time for optical
pulses to propagate through the fibre with different central
wavelengths. The GVD is obtained by dividing the group velocity
differences by the wavelength shift. The nonlinear parameter
$\gamma$ is determined using $\gamma=2\pi n_2/\lambda A_{eff}$,
where the nonlinear-index coefficient
$n_2\sim$$2.5\times10^{-20}$m$^{2}$/W is used for silica.
\begin{table}[h]
\centering \caption{Parameters of the fibres used in the
experiment.}
\begin{tabular}{@{}llllll} \br
  &$A_{eff}$&$L_{eff}$&$G/P_i$(linear region)&$\gamma$&$\beta_2$\\\hline
  LEAF & 72 $\mu$m$^2$&12.8 \ km &10.5 W$^{-1}$ & 1.4 W$^{-1}$km$^{-1}$ & -5.29 ps$^2$/km \\
  HNLF & 11.7 $\mu$m$^2$ & \ 1.64 km &11.1 W$^{-1}$ & 8.7 W$^{-1}$km$^{-1}$ & \ 0.08 ps$^2$/km  \\
 \br
\end{tabular}

\end{table}

As one of the most widely used low-cost standard single-mode fibre,
LEAF fibre offers great compatibility and would substantially lower
the cost of SBS slow light devices, rendering it the first candidate
for our experiment. It turns out, however, the zero-dispersion
wavelength for the LEAF is not shifted to the transparency window,
resulting in a large anomalous dispersion $\beta_2$ at $\sim$1.55
$\mu$m, which degrades the SBS slow light performance, as shown
below.

The SBS gain as a function of $P_i$ is measured for the LEAF fibre
and is shown in figure 2a. As $P_i$ is increased from zero to $0.8\
$W, $G$ first scales linearly with $P_i$, with a slope of 10.5
W$^{-1}$, and then starts to saturate and deviate from linear growth
at $G\sim$4, corresponding to an input power $P_i\sim$0.4 W (figure
2a). The early saturation of the SBS gain limits the SBS slow light
delay.

The saturation behavior shown in figure 2a could be the result of
pump depletion due to the SBS process, as we mentioned previously.
To rule out this effect, we note the following. First, the saturated
value of the gain is $\sim$4, which is much smaller than the
threshold gain $G_{{th}}=10$, indicating that saturation is not
induced by spontaneous Brillouin amplification. Second, the
amplified signal power $P_s$ is only 0.5 mW, which is small in
comparison to the input pump power 0.75 W, and is hence much too
small to deplete the pump. Moreover, the total power $P_o$
transmitted through the fibre in the direction of the pump beam
grows linearly as $P_i$ is increased from 0 to 0.8 W, as shown in
figure 2b, which shows directly that the total transmitted power is
not depleted.

\begin{figure}[tb]\label{saturation}
\flushright\includegraphics[width=11cm]{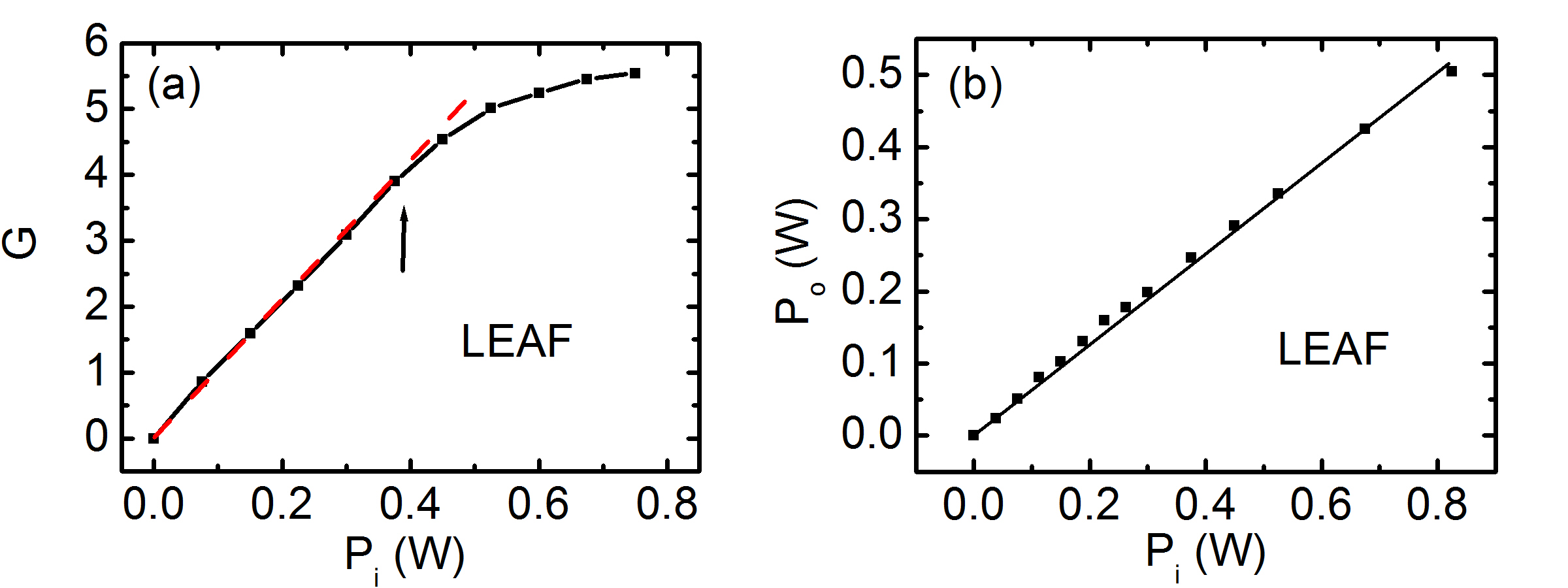}
 \caption{Early saturation
of the SBS gain. (a) The SBS gain $G$ as a function of input pump
power $P_i$. Saturation is observed around $G\approx4$ (vertical
arrow), corresponding to an input pump power of $\sim$0.4 W; (b)
Total power $P_o$ transmitted through the fibre in the pump beam
direction (not spectrally resolved) as a function of the input pump
power $P_i$, indicating high fibre transparency. In the experiment,
a weak continuous-wave beam with an input power of 2 $\mu$W is used
as the signal beam.}
\end{figure}

\begin{figure}[htb]\label{leaf}
\flushright\includegraphics[width=12.5cm]{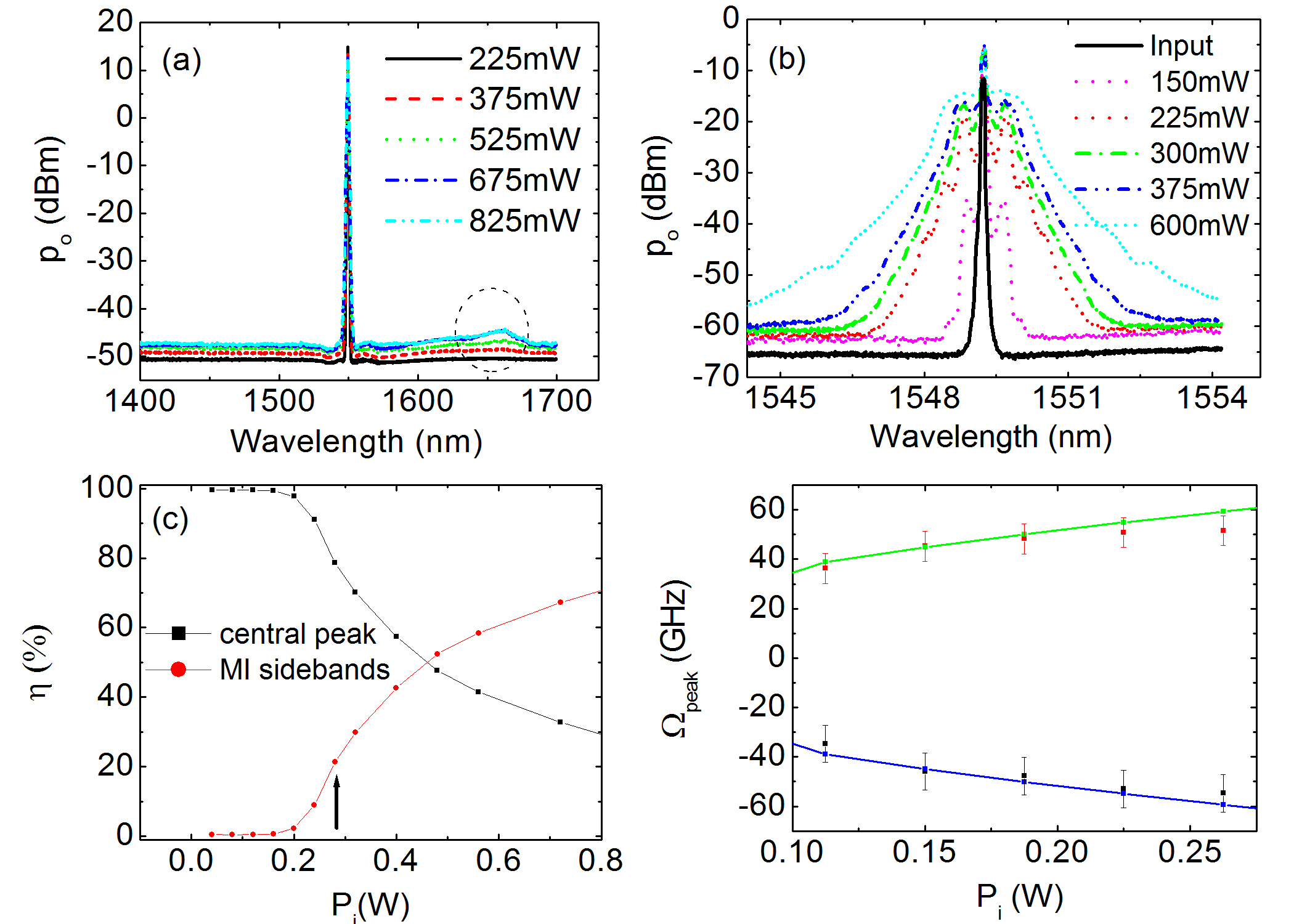} \caption{Modulation
instability in the LEAF fibre. (a) Output power spectral density
$p_{o}$ of the pump laser (span of 300 nm) with increasing input
power $P_i$. The circle indicates the Raman peak. (b) Output power
density spectra $p_{o}$ of the pump laser (span of 10 nm) with
increased input power $P_i$. The input spectrum of the pump laser
(at 40 mW) is also shown. (c) Percentage ($\eta$) of power
distributed in the central peak (black square) and in the MI
sidebands (red circle) for the output pump beam, as functions of the
input power $P_{i}$. The arrow indicates threshold at $\sim$0.3 W.
(d) Experimental data (point) and theoretical prediction (line) of
the frequency shift $\Omega_{peak}$ of the MI side lobe peaks as a
function of input power $P_{i}$. }
\end{figure}

To explain the early SBS gain saturation at $G\sim$4, we examine the
transmitted pump spectrum $p_o$ passing through the 20-km-long LEAF
fibre  (Anritsu model MS9710B optical spectrum analyzer). As $P_i$
is increased from zero to 0.8 W, no significant Raman gain is
observed (\fref{leaf}a). On the other hand, in spectral span of 10
nm, symmetric side lobe structures emerge and grow quickly as $P_i$
increases (\fref{leaf}b). At high $P_i$, the spectrum of the pump is
flattened as a result of the emergence of secondary sidebands, and
the power is spread into a broad frequency span of $\sim$200 GHz.
Notice that the pump power transferred to the side lobes is no
longer on resonance with the signal beam in the SBS interaction and
therefore does not contribute to the SBS gain process.

To determine the amount of power that is transferred to the side
lobes, we integrate the power spectral density $p_o$ of the output
pump beam and calculate the distracted power as percentages $\eta$
of the total power. As shown in \fref{leaf}c, a considerable
proportion ($>20\%$) of the pump power is transferred to the
sidebands when the input power exceeds $P_i\sim$0.3 W. We define
this point as the threshold for the MI process and the corresponding
input pump power as the threshold power $P_{th}^{MI}$ in the LEAF
fibre. The threshold gain $G_{th}^{MI}$ is obtained by \eref{gainMI}
to be $\sim$10. Note that $P_{th}^{MI}$ is close to the location
where the early saturation of SBS gain occurs. We also compare the
measured frequency shifts $\Omega_{peak}$ of the MI sidelobe peaks
with the theoretical prediction using equation \eref{peak} and
obtain good agreement, as shown in \fref{leaf}d. These observations
lead us to conclude that the strong saturation of the SBS gain is
caused by the MI-induced pump broadening. In conclusion, we find
that MI dominates over SBS beyond a threshold pump power of
$P^{MI}_{{th}}=0.3$ W in the LEAF fibre. In a broadband system where
$\Gamma=9.2$ GHz, the threshold pump power corresponds to an SBS
gain $G\sim$3.2, which leads to a limit of 29 GHz or 126
dB$\cdot$GHz on the SBS gain-bandwidth product.

\begin{figure}[htb]\label{no_MI}
\centering\includegraphics[width=13cm]{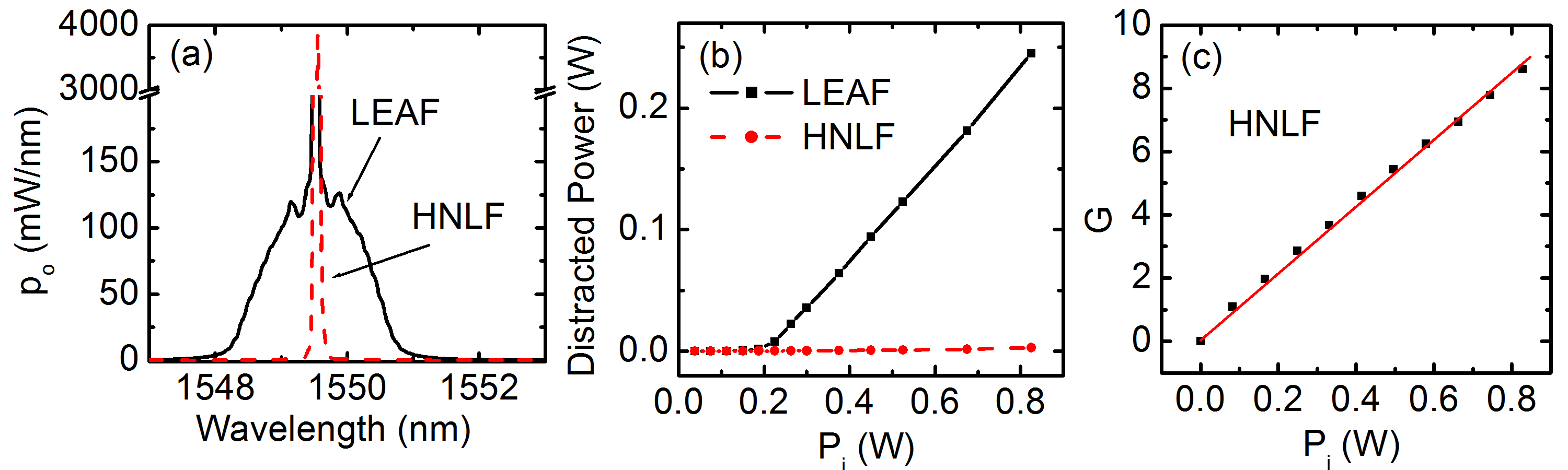} \caption{Suppression of
modulation instability in HNLF. (a) Output power spectrum $p_{o}$ of
HNLF (red dash) and LEAF (black solid) at an input power of 0.8 W;
(b) MI distracted pump power as a function of input power $P_{i}$ in
HNLF (red dot) and LEAF (black square).(c) SBS gain $G$ in HNLF as a
function of the input power $P_{i}$. }
\end{figure}
We next turn to the dispersion-shifted HNLF fibre, which has a small
and normal dispersion at 1.55 $\mu$m. Because it is widely known
that the MI is suppressed due to phase matching constraints in
normally dispersive materials
\cite{NonlinearFiberOptics,MIinFiber,MIpulse}, we expect to see
suppression of the MI and improvement in the gain-bandwidth product
for the broadband SBS slow light with the HNLF fibre. As shown in
figure 4, it is found that the MI-induced limit on the SBS
gain-bandwidth product is indeed removed. With the same 9.2 GHz
broadband pump input, the transmitted pump spectrum through the HNLF
shows no significant MI peaks (figure 4a). The power converted into
off-resonant frequencies is negligible (figure 4b). As expected,
early saturation of the SBS gain in the HNLF does not appear (figure
4c), resulting in a larger SBS gain-bandwidth product of 344
dB$\cdot$GHz, limited by our available pump power of 0.82 W. The
result further confirms that MI induced pump broadening is the
origin of the early saturation of SBS gain in the LEAF fibre.

\section{Discussion}

In this section, we compare the relative importance of the two
effects that compete with SBS in our broadband SBS slow light
system. Following Olsson \textit{et al.}'s approach, we compare the
pump power $P_i$ requirements for the broadband SBS, SRS, and MI
processes in the LEAF fibre. Figure 5 shows the input pump power
$P_i$ required to obtain a threshold gain of 10 as a function of the
bandwidth $\Gamma$. The threshold pump power for the SBS process is
obtained from equation \eref{accurategain} taking $G$ =10 and $g_0 =
1.06 \times 10^{-11}$ m/W (obtained from figure 2a), giving
\begin{equation}
P_i=G\frac{\Gamma}{2\Gamma_B}\frac{A_{eff}}{g_0L_{eff}}\arctan^{-1}(\Gamma/\Gamma_B).
\end{equation}
Note that we assume that a weak probe beam is used so that the
probe-induced SBS saturation does not appear.

The SRS gain is given by \cite{NonlinearFiberOptics}
\begin{equation}\label{SRS}
G_{SRS}=2g_RP_iL_{eff}/A_{eff},
\end{equation}
where $g_R=1.26\times10^{-14}$ m/W \cite{olsson1987}. The threshold
power required for an SRS gain of 10 is obtained by
$P_{i}=10A_{eff}/(2g_RL_{eff})=2.23$ W. The pump power required for
a MI gain $G_{MI}\sim$10 is found to be $\sim$0.3 W, as mentioned
previously.
\begin{figure}[htb]\label{compare}
\centering\includegraphics[width=6cm]{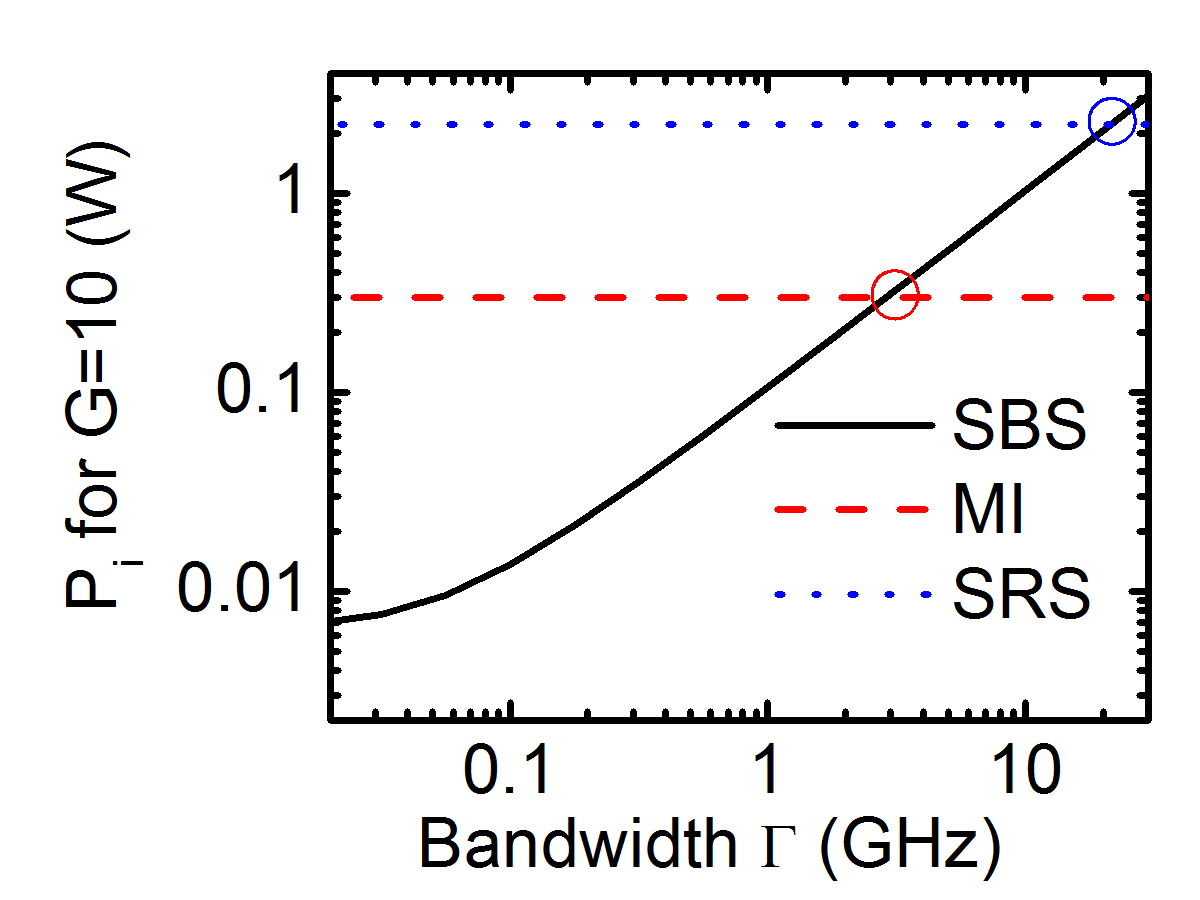} \caption{Pump power $P_i$
required for a $G$ of 10 versus the bandwidth $\Gamma$.}
\end{figure}

As shown in figure 5, beyond the bandwidth of $\sim$3.2 GHz, the
LEAF-fibre system hits the MI threshold before it saturates the SBS
gain. As a result, MI sets a limit to the SBS gain-bandwidth
product. The gain-bandwidth product limit is $29$ GHz, or $126$
dB$\cdot$GHz using G=10 in the LEAF fibre. Figure 5 also shows that
the SRS becomes more efficient than the SBS when the bandwidth goes
beyond $\sim$22 GHz, and limits the SBS gain-bandwidth product up to
220 GHz or 955 dB$\cdot$GHz. The result is consistent with what
Olsson \textit{et al.} predict \cite{olsson1987}. In the LEAF fibre,
the SBS gain-bandwidth product is restricted by the tighter
MI-induced limit. However, in normally dispersive fibres where the
MI is suppressed, SRS becomes the main limitation on the SBS
gain-bandwidth product.

In conclusion, we found that in broadband SBS systems, MI dominates
in the high pump power region and sets a limit to the SBS gain
bandwidth product, which is 126 dB$\cdot$GHz in the LEAF fibre. The
SBS suppression and SBS slow light performance is improved in
normally dispersive materials.

\section*{Acknowledgements}

We gratefully acknowledges the financial  support of the DARPA
Defense Sciences Office Slow Light project.

\end{document}